\documentclass[aps,pre,groupedaddress,showpacs,showkeys,twocolumn,amsmath,amsfonts,amssymb]{revtex4-1}
\usepackage{graphicx}

\usepackage{xcolor}
\usepackage{ulem}
\colorlet{MyGreen}{black!50!green}
\colorlet{MyOrange}{red!70!yellow}

\begin{document}

%\preprint{}

\title{A microcanonical entropy correcting finite-size effects in small systems}


\author{Roberto Franzosi}
\email[]{roberto.franzosi@ino.it}
%\thanks{}
\affiliation{QSTAR \& CNR - Istituto Nazionale di Ottica, Largo Enrico Fermi 2, I-50125 Firenze, Italy}


\date{\today}

\begin{abstract}

In a recent paper [Franzosi, Physica A {\bf 494}, 302 (2018)], we have suggested to use of the surface entropy, namely the logarithm of the area of a hypersurface of constant energy in the phase space, as an expression for the thermodynamic microcanonical entropy, in place of the standard definition usually known as Boltzmann entropy. In the present manuscript, we have tested the surface entropy on the Fermi-Pasta-Ulam model for which we have computed the caloric equations that derive from both the Boltzmann entropy and the surface entropy. The results achieved clearly show that in the case of the Boltzmann entropy there is a strong dependence of the caloric equation from the system size, whereas in the case of the surface entropy there is no such dependence.
We infer that the issues that one encounters when
the Boltzmann entropy is used in the statistical
description of small systems could be a clue of a deeper defect of this entropy that derives from its
basic definition.
Furthermore, we show that the surface entropy is well founded from a mathematical point of view, and we show that it is the only admissible entropy definition, for an isolated and finite system with a given energy, which is consistent with the postulate of equal a-priori probability.
 
\end{abstract}


\maketitle

%\section{}
%\subsection{}
%\subsubsection{}

\section{Introduction}
The thermodynamics of an isolated system is described by the microcanonical
ensemble, in which the averages of all the physical quantities are derived
by the entropy. Thus, the starting point for the statistical description
of any isolated system is the definition for the entropy.
In the case of a classical Hamiltonian system, there are two
accepted choices: the Boltzmann entropy and the Gibbs entropy.
Recently, a third entropy definition has been proposed in
\cite{Franzosi_PhysicaA18}, there \textit{the surface entropy} has
been defined similarly to
the Boltzmann entropy but without a limit for a spread of
energy that vanishes.
In Ref. \cite{Franzosi_PhysicaA18} we have proved that in the
limit of a large number of degrees of freedom, the surface entropy
predicts the same results as the Boltzmann entropy, included the
possibility to observe negative temperatures.
In the present work we
show that, on the contrary, in the case of small systems these
entropies can be inequivalent.
Here we show, in fact, that the surface entropy has properties
which make it an attractive definition for small systems.
In particular, in Sec. \ref{derivation} is
reported the main result: We derive the surface entropy
from first principles, we show that this definition is
the only admissible for a classical isolated system with
a given energy, which is consistent with the postulate of
equal a-priori probability. Before this, in Sec. \ref{evidences},
we report results of simulations on a Fermi-Pasta-Ulam non-linearly
coupled oscillator chain, that show that in the case of the surface
entropy there is not an appreciable dependence of the caloric equation
from the system size, whereas, in the case of the Boltzmann entropy
there is an evident dependence.

Let $H(x)$ be a classical Hamiltonian describing an autonomous
many-body system of $n$ interacting particles in $d$ spatial dimensions.
Let the coordinates and canonical momenta $(q_1\ldots, p_1 ,\ldots)$ be
represented as $N$-components vectors $x\in \mathbb{R}^{N}$, with $N=2nd$.
Let 
$
M_E = \left\{x\in \mathbb{R}^{N} | H(x) \leq E \right\}
%\label{ME}
$
be the set of phase-space states with total energy less than or
equal to $E$.
The Gibbs entropy for this system is
\begin{equation}
S_G (E) = \kappa_B \ln \Omega(E) \, ,
\label{gibbs}
\end{equation}
where $\kappa_B$ is the Boltzmann constant and
\begin{equation}
\Omega(E) = \dfrac{1}{h^{nd}} \int d^N x \Theta(E-H(x))
\label{OmegaE}
\end{equation}
is the number of states with energy below $E$,  $h$ is the Planck constant and
$\Theta$ is the Heaviside function.
Conversely, the Boltzmann entropy pertains to the energy-level sets 
$
\Sigma_E = \left\{x\in \mathbb{R}^{N} | H(x) = E \right\}
%\label{SigmaE}
$
and is given in terms of $\omega(E) = \partial \Omega/\partial E$, according to
\begin{equation}
S_B (E) = \kappa_B \ln \left(\omega(E)\Delta \right) \, ,
\label{boltzmann}
\end{equation}
where the constant $\Delta$ has dimension of energy and makes
the argument of the logarithm dimensionless, and
\begin{equation}
\omega(E) = \dfrac{1}{h^{nd}} \int d^N x \delta(E-H(x)) \, ,
\label{omegaE}
\end{equation}
is expressed in terms of the Dirac $\delta$ function. Remarkably, in the case
of smooth level sets $\Sigma_E$, $\omega(E)$
can be cast in the following form 
\cite{RughPRL97,Franzosi_JSP11,Franzosi_PRE12}
\begin{equation}
\omega(E) =  \dfrac{1}{h^{nd}} 
\int_{\Sigma_E} \dfrac{m^{N-1}(\Sigma_E)}{\Vert\nabla H(x) \Vert}  \, ,
\label{omegaEdiff}
\end{equation}
where $m^{N-1}(\Sigma_E)$ is the metric induced from $\mathbb{R}^N$ on the
hypersurface $\Sigma_E$ and $\Vert\nabla H(x) \Vert$ is the norm of the gradient
of $H$ at $x$. 
Finally,
in Ref. \cite{Franzosi_PhysicaA18} we have proposed the
definition of the surface entropy
\begin{equation}
S (E) = \kappa_B \ln \left( \sigma(E) \Lambda^{1/2} \right) \, ,
\label{enew}
\end{equation}
where the constant $\Lambda$ has the dimension of action and makes
the argument of the logarithm dimensionless and
\begin{equation}
\sigma(E) =\dfrac{1}{h^{nd}}  \int_{\Sigma_E} m^{N-1}(\Sigma_E)  \, ,
\label{sigmaEdiff}
\end{equation}
is the measure of $\Sigma_E$.
Very recently, the long-standing debate on which one is the correct
entropy definition, between the Boltzmann's and the Gibb's one, has
became a topical one after the papers of Refs. 
\cite{Dunkel2013,Hilbert_PRE_2014,Hanggi_15}, in which it has been
argued that the Gibbs entropy yields a consistent thermodynamics,
whereas the Boltzmann entropy would have some consistency issues.
This fact, indeed, would have dramatic consequences about the
foundations of statistical mechanics. For instance, the
notion of negative temperatures is a well-founded concept in
the Boltzmann description, whereas it is forbidden in the
Gibbs description.
On the contrary, in Refs. \cite{Buonsante_AoP_2016, Buonsante_2015},
we have shown that the Boltzmann entropy provides a consistent
description of the microcanonical thermostatistics of  \textit{macroscopic
systems}, and, consequently, negative temperatures are a well-posed
concept.
For instance, in Ref. \cite{Buonsante_AoP_2016}  we show that, in the
thermodynamic limit, the Gibbs and Boltzmann temperatures do coincide
when the latter is positive whereas the inverse Gibbs temperature is
identically null in the region where Boltzmann provides negative values
for the temperature. This means that in correspondence of the energies
where $\beta_B$ is negative, none microcanonical description based on
the Gibbs entropy is possible, albeit the region of energies corresponding
to $\beta_B < 0$ is absolutely accessible to microscopic dynamics.

Nonetheless, the relation between the total energy and the
inverse temperature, $E(\beta)$, for instance in the simple
case of an isolated ideal
gas system, derived with the standard Boltzmann entropy, is not
an extensive quantity.
Although this fact does not represent a problem in the
case of macroscopic systems, since the correction to the
extensive behaviour is of the order of $1/(nd)$, it
may be an issue in the case of small systems since the
measure of temperature (and other thermodynamic quantities)
in such systems could be inaccurate.
In order to overcome such issues, we have recently proposed the
surface entropy \eqref{enew},
that reproduces the same results as the Boltzmann entropy
for systems with a macroscopic number of particles but it
predicts the correct extensivity for the total energy derived
by the caloric equation,
in the case of small systems \cite{Franzosi_PhysicaA18}.
Therefore, in our opinion,
the definition \eqref{enew} represents a valid candidate
for the solution of the problem regarding the correct
statistical description of isolated finite systems.

Actually, a check of consistency of the surface entropy
is lacking and the purpose of the present manuscript is,
indeed, to provide a mathematical foundation of this entropy
definition. Before starting such a discussion, we report the
results of numerical simulations.

\section{Paradigmatic evidences \label{evidences}}
As an example, we consider the celebrated Fermi-Pasta-Ulam (FPU)
$\beta$-model \cite{FPU}, described by the Hamiltonian
\begin{equation}
H = \sum^n_{j=1} 
\left[
\dfrac{1}{2}p^2_j + \dfrac{1}{2} (q_j - q_{j+1})^2 + \dfrac{\mu}{4}
 (q_j - q_{j+1})^4
\right] \, ,
\label{FPU-H}
\end{equation}
where we have taken periodic boundary conditions.
The FPU is a nonlinear chaotic system and it has been largely
used as a toy model since the first numerical experiment
by Fermi and co-authors.
This apparently simple system has unveiled intriguing behaviours as to
the dynamical instability \cite{Pettini_Chaos05}
and this is one of the reasons for its interest.
Within the present study on the surface entropy, such a model exhibits
two-fold convenience. Firstly, in this case, the statistical
ensembles equivalence holds true and thermodynamic quantities
derived in the canonical and in the microcanonical ensemble
will agree in the thermodynamic limit. Secondly, the kinetic
energy has the standard form and this permits to perform
some analytical calculations \cite{Pearson85}.  
In addition, Hamiltonian \eqref{FPU-H} is unbounded from above and
therefore the thermodynamics derived from the Boltzmann and Gibbs
entropy are consistent \cite{Buonsante_AoP_2016}, at least in
the case of macroscopic number of degrees of freedom.
In the following, we report on a numerical analysis performed
on the FPU model in which we have measured appropriate quantities
along the dynamics generated by Hamiltonian \eqref{FPU-H},
in order to derive the caloric equation either from the Boltzmann
entropy and from the surface entropy. In order to test the reliability
of the two entropy definitions, the simulations have been performed
for several system sizes.

As far as the standard Boltzmann entropy, the form of Hamiltonian
\eqref{FPU-H} allows using the Laplace-transform technique
\cite{Pearson85} in order to measure  
the Boltzmann inverse temperature
\begin{equation}
\beta_B = \left(\dfrac{n}{2} -1 \right)
\langle K^{-1} \rangle_{\mu c} \, ,
\label{Pearson}
\end{equation}
where $K$ is the kinetic energy of \eqref{FPU-H} and 
where we indicate with $\langle \cdot \rangle_{\mu c}$ the time average of a 
given observable, taken over the trajectories corresponding to several
initial conditions.
In Ref. \cite{Franzosi_PhysicaA18}, we have shown that in the case
of the surface entropy it results
\begin{equation}
\beta_s = 
\dfrac{\partial \sigma/\partial E}{\sigma} =
\dfrac{(\partial \sigma / \partial E)/\omega}{\sigma/\omega} =
\dfrac{\langle \nabla \left( \frac{\nabla H}{\Vert \nabla H \Vert} 
\right) \rangle_{\mu c}}
{\langle \Vert \nabla H \Vert \rangle_{\mu c}} \, ,
\label{betaSurf}
\end{equation}
thus $\beta_s$ is derived from the ratio between the time averages of
$
\nabla \left( {\nabla H}/{\Vert \nabla H \Vert} 
\right)
$ 
and $\Vert \nabla H \Vert$, taken over the trajectories in the 
phase space, where
$\nabla$ is the gradient in the $N$-dimensional phase space
and $\Vert \cdot \Vert$ is the standard Euclidean norm.
In addition to Eqs. (\ref{Pearson}) and (\ref{betaSurf}) for
$\beta_B$ and $\beta_s$,
by resorting to the equipartition theorem we get 
a third alternative definition of temperature
\footnote{It is worth emphasizing that for a system with energy
unbounded from above, as in the case of the FPU model, $\beta_{et}$
coincides with $\beta_G$ \cite{Buonsante_AoP_2016} in the thermodynamic
limit.}
\begin{equation}
\beta_{et} = \dfrac{n}{2\langle K \rangle_{\mu c}} \, .
\label{betaet}
\end{equation}

The results of our simulations 
show that the mapping between temperature and energy
derived from the latter expression provide a curve that can be
assumed as the reference when the system size is varied.
All the simulations have been performed with $\mu=0.1$.
Note that, the relevant energy region is that of high energies,
definitely above the strong stochasticity threshold, namely in the
region where the system is ergodic \cite{chaos_15,PhysRevE.78.046205}.
\begin{figure}[h]
 \includegraphics[width=7.cm]{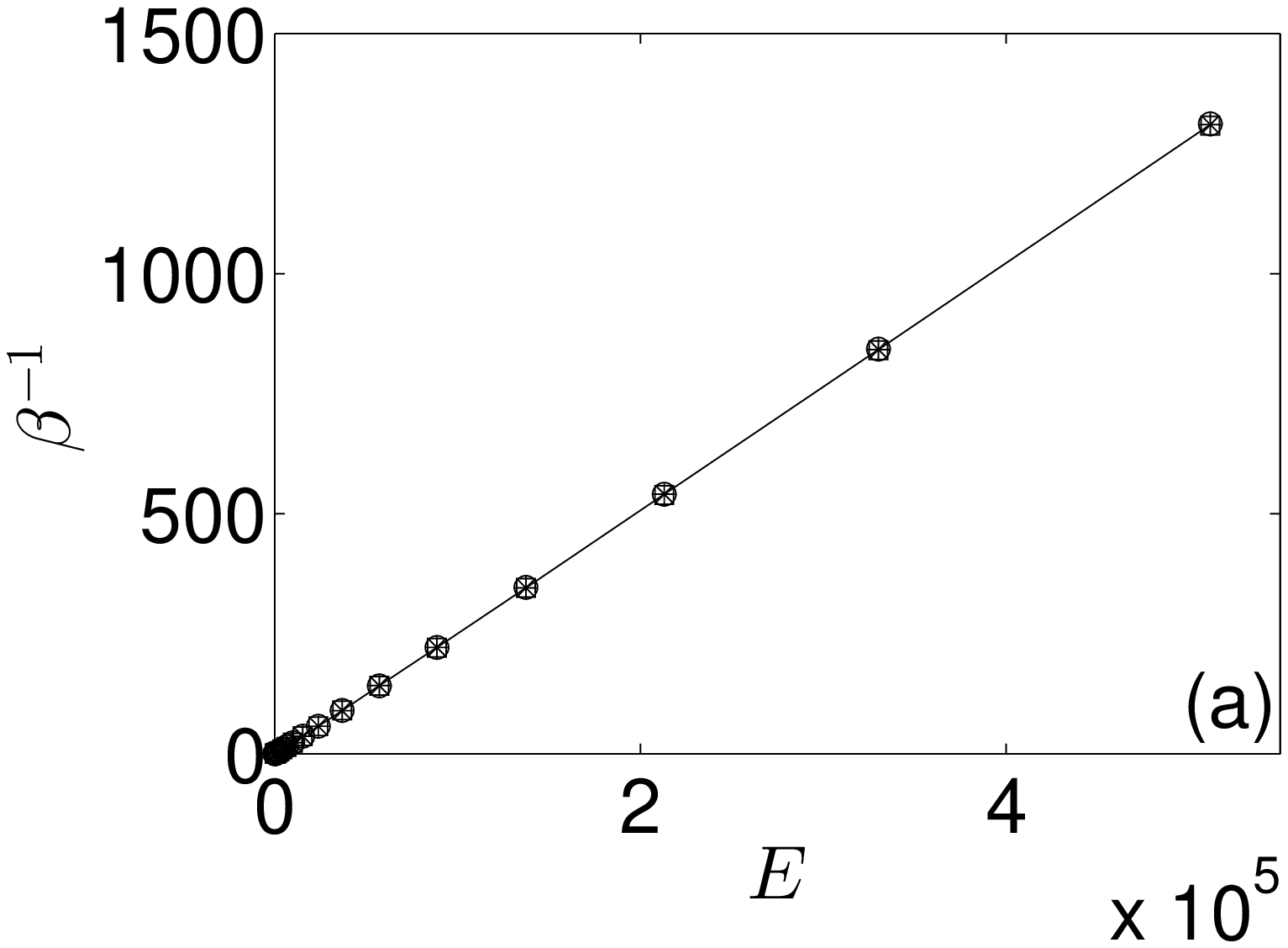}
 \includegraphics[width=7.cm]{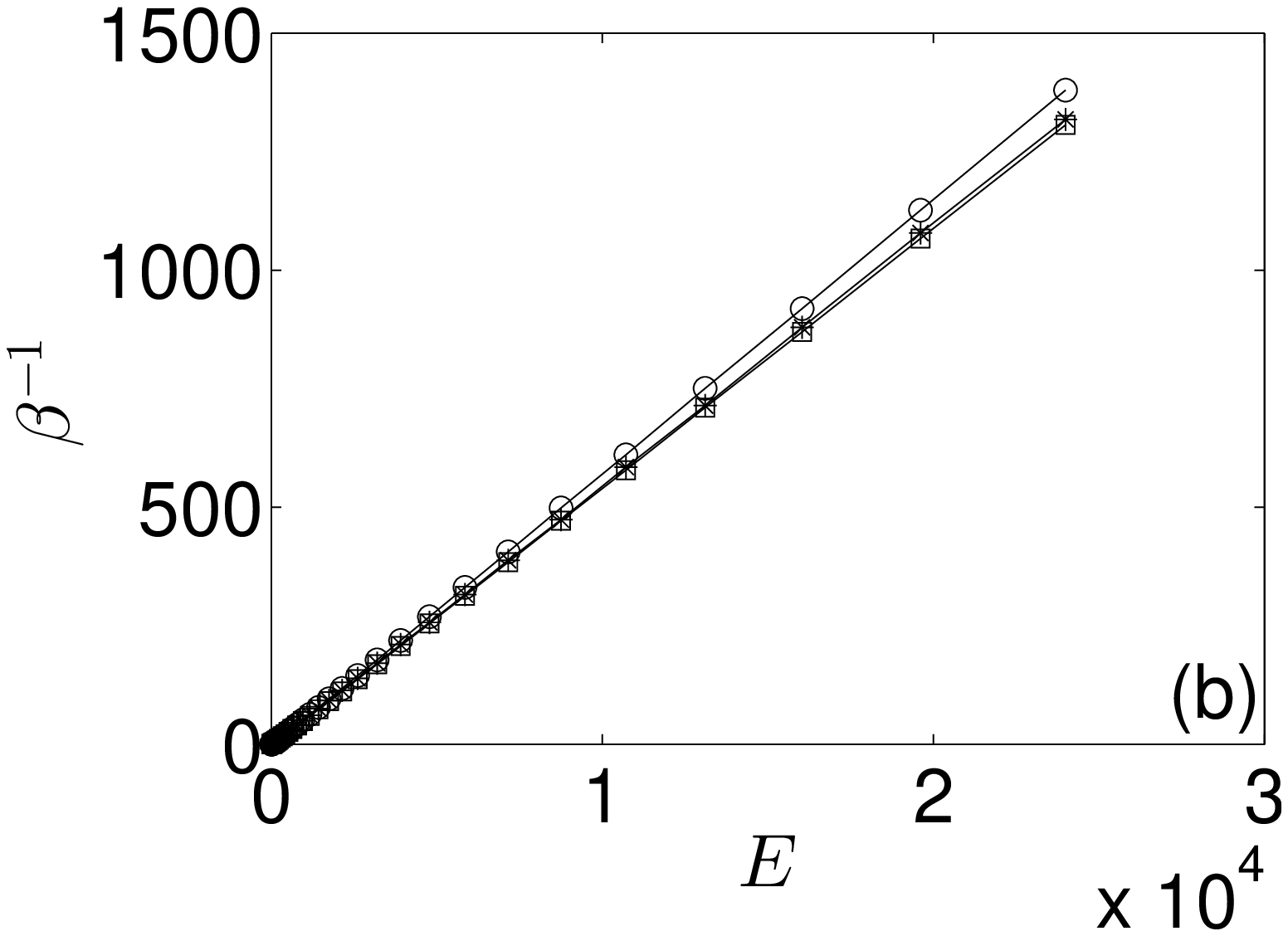}
 \includegraphics[width=7.cm]{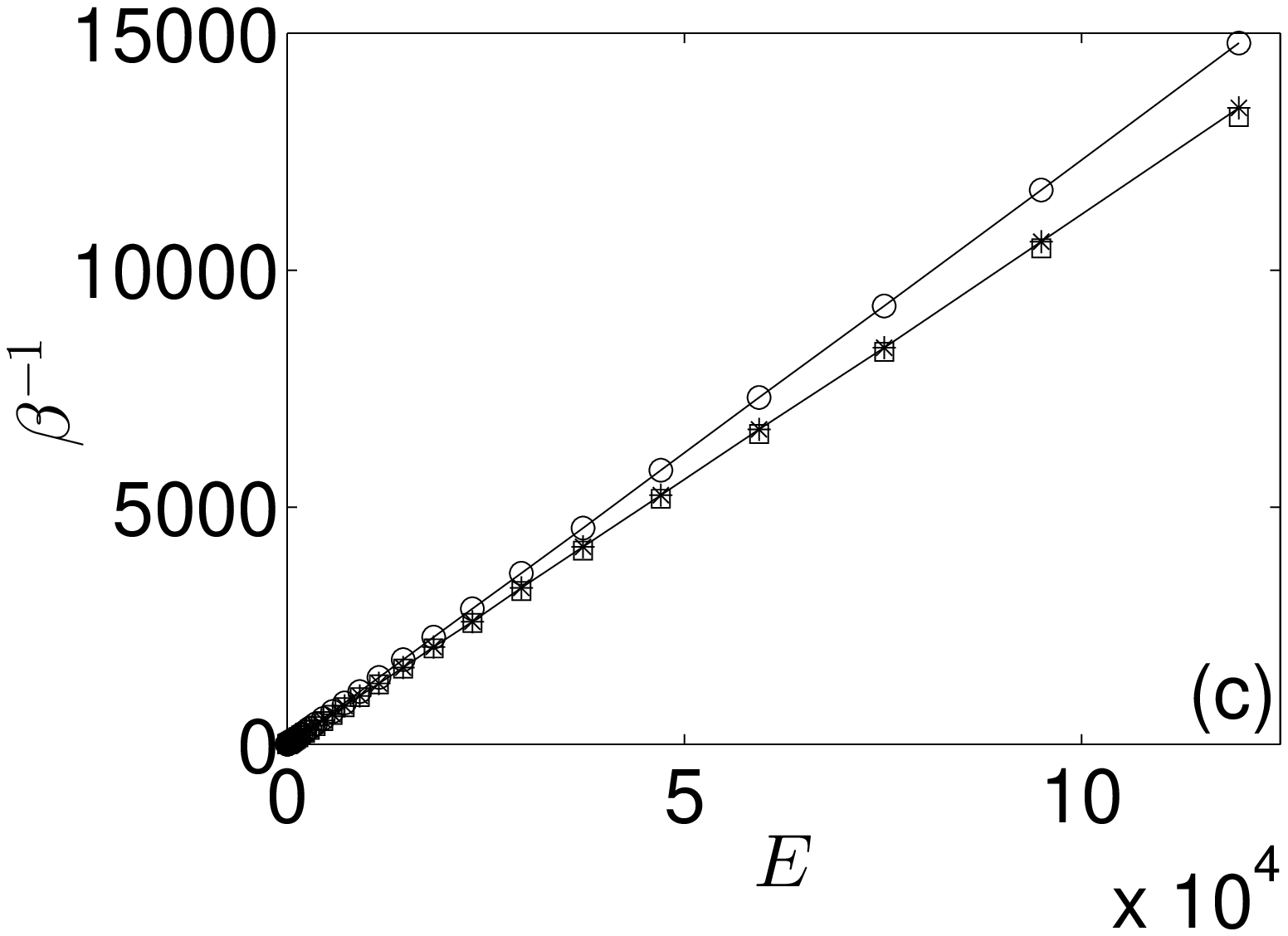}
\caption{The figure compares the $\beta^{-1}$ derived from the three
different definitions \eqref{Pearson}, \eqref{betaSurf} and \eqref{betaet}
and measured for several values of the total density energy $E$ 
for different particle numbers: (a) $n=512$, (b) $n=24$, (c) $n=12$.
We report the measured values for $\beta_s^{-1}$ 
using stars ($\star$),
the measured values for $\beta^{-1}_{et}$ using open squares ($\Box$)
and the measured values for $\beta^{-1}_{B}$ using open circles ($\circ$).
\label{Fig_1_new}}
\end{figure} 
In Fig. \ref{Fig_1_new} we report the comparison of the curves 
$\beta^{-1}(E)$ derived from the three
different definitions \eqref{Pearson}, \eqref{betaSurf} and \eqref{betaet}
and measured for several values of the total density energy $E$ and 
some system sizes.
Already in the case of a system size of
$512$ particles, the the inverse temperatures derived from the three
entropy definitions agree in an excellent way.
Remarkably, also in the case of a very small system size, as in the case
reported in Fig. \ref{Fig_1_new} panel (c), the two definitions
\eqref{betaSurf} and
\eqref{betaet} have a very good accordance.
Nevertheless, the smaller the system size, the higher the discrepancy 
between $\beta^{-1}_B$  and the measured values of $\beta_s^{-1}$ and
$\beta^{-1}_{et}$.
\begin{figure}[h]
 \includegraphics[width=7.cm]{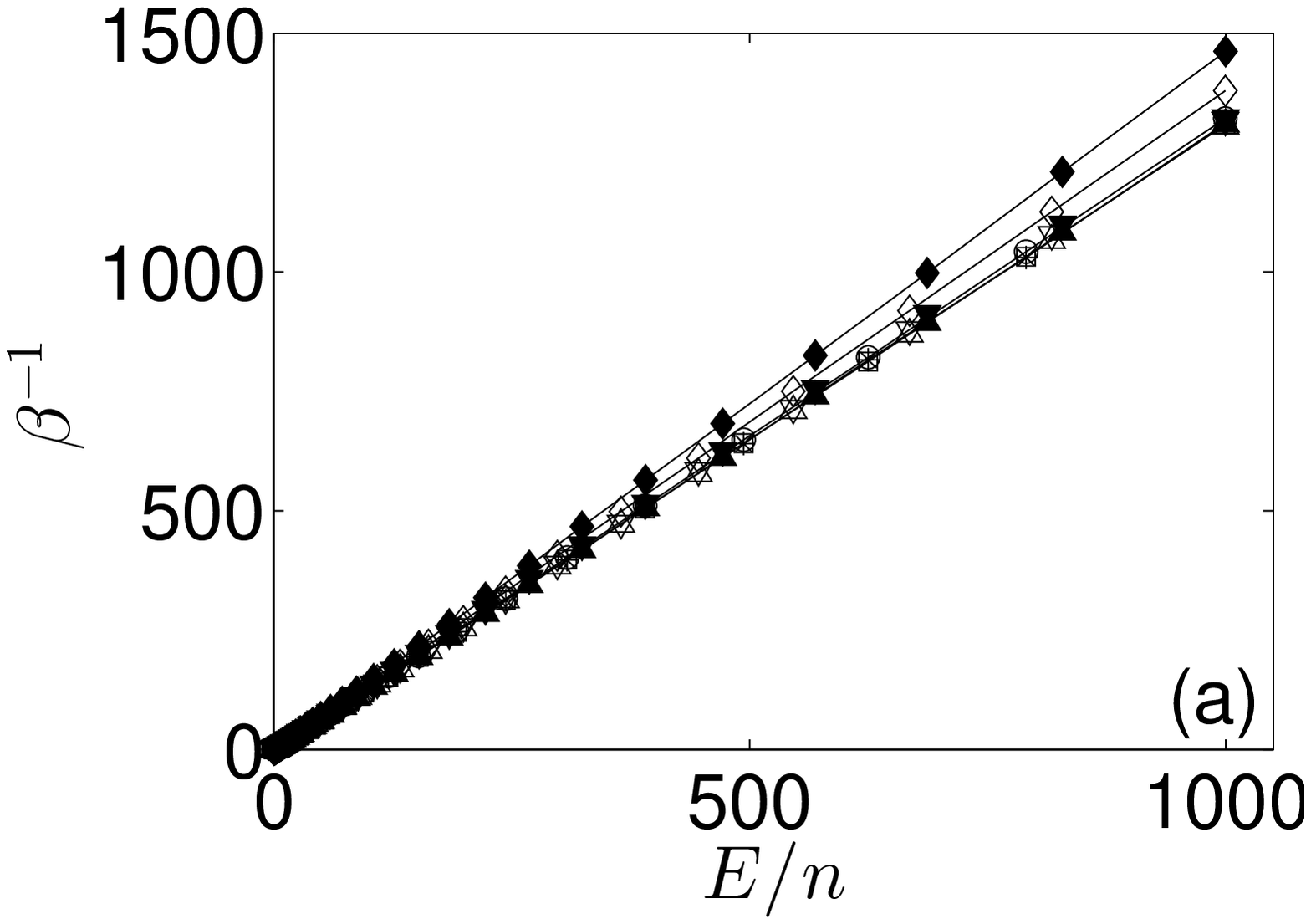}
 \includegraphics[width=7.cm]{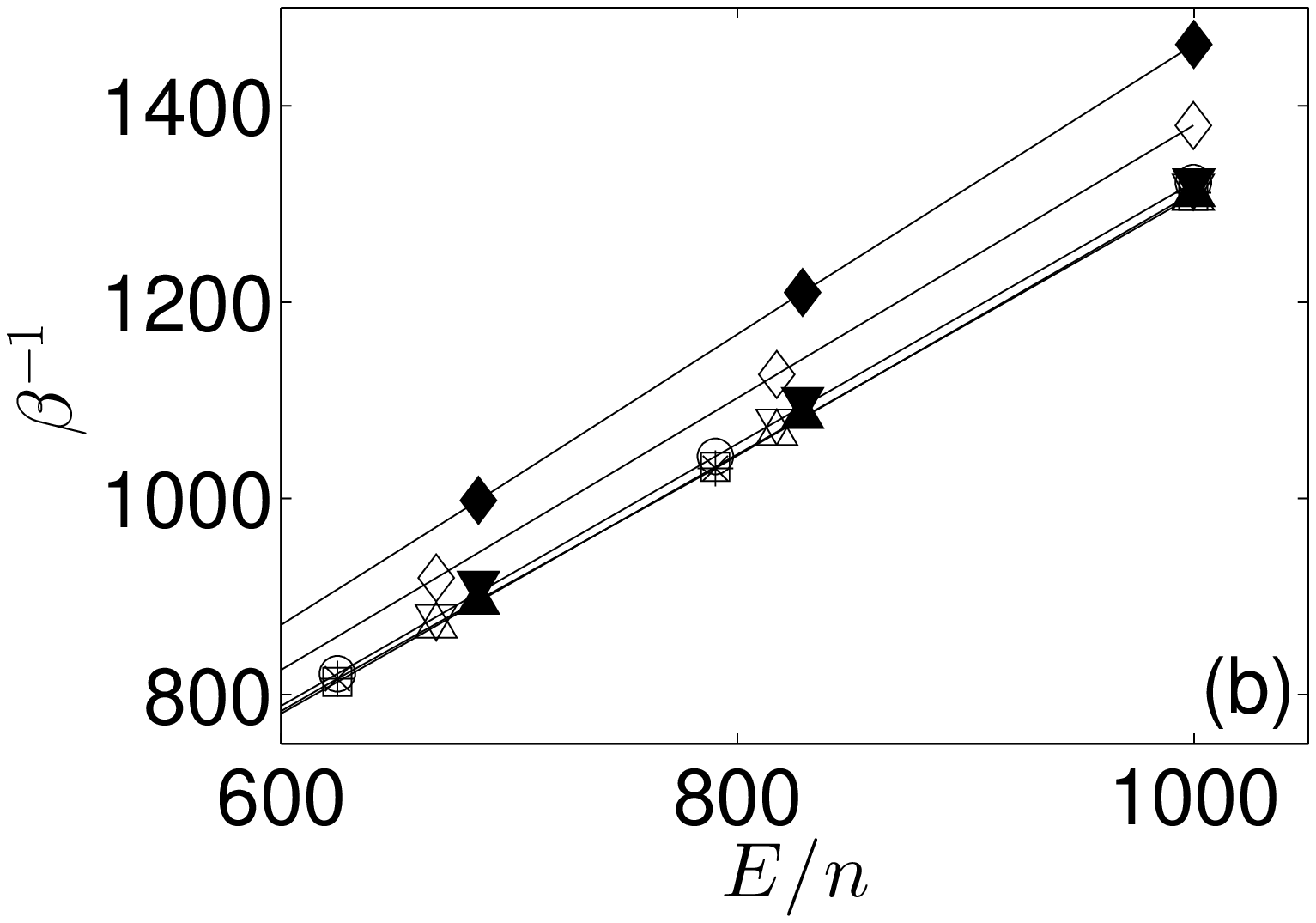}
\caption{The figure compares the $\beta^{-1}$ derived from the three
different definitions \eqref{Pearson}, \eqref{betaSurf} and \eqref{betaet}
and measured for several values of the total density energy $E/n$ and
three system sizes.
In panel (a), with stars ($\star$) we report the measured values
for $\beta^{-1}_s$ in the case
$n=128$,  
with open squares ($\Box$) the measured values for
$\beta^{-1}_{et}$ in the case
$n=128$ and
with open circles ($\circ$) the measured values for
$\beta^{-1}_{B}$ in the case
$n=128$.
With open down triangles ($\triangledown$) we report the measured values
for $\beta^{-1}_s$ in the case
$n=24$,  
with open up triangles ($\triangle$) the measured values for
$\beta^{-1}_{et}$ in the case
$n=24$ and
with open diamonds ($\lozenge$) the measured values for
$\beta^{-1}_{B}$ in the case
$n=24$.
With full down triangles ($\blacktriangledown$) we report the measured values
for $\beta^{-1}_s$ in the case
$n=12$,  
with full up triangles ($\blacktriangle$) the measured values for
$\beta^{-1}_{et}$ in the case
$n=12$ and
with full diamonds ($\blacklozenge$) the measured values for
$\beta^{-1}_{B}$ in the case
$n=12$.
Panel (b), is a magnification of a small region of panel (a).
\label{Fig_2_new}}
\end{figure} 
In Fig. \ref{Fig_2_new} panel (a) we compare the curves $\beta^{-1}(E/n)$
derived from the three
definitions \eqref{Pearson}, \eqref{betaSurf} and \eqref{betaet} and
measured for several values of the density of the total energy $E/n$
and for three system sizes. Fig. \ref{Fig_2_new} panel (b) is a
magnification of a small region in 
Fig. \ref{Fig_2_new} (a).
Both the figures display the features described
above: $\beta^{-1}$ derived according to the two
definitions \eqref{betaSurf} and \eqref{betaet} show a very good
accordance one another for large as well as small system size, 
and the smaller the system size, the higher the deviation
between $\beta^{-1}_B$  and $\beta_s^{-1}$ and
$\beta^{-1}_{et}$.

This analysis shows that the thermodynamic quantities derived
from Boltzmann entropy in the case of a small system can be
affected by relevant deviations from the asymptotic behaviour.
On the contrary, the same does not happen when using the
surface entropy. This seems to indicate that the microstatistical
description of ``intrinsically small'' systems
(nanoparticles, proteins, cellular systems) could be
profitably improved by resorting to the surface entropy
instead of the Boltzmann entropy.

Furthermore, in our opinion, the deviation of $\beta^{-1}_B$
from $\beta^{-1}_{et}$ observed when the system size is
decreased can bring to a serious logical issue. In fact,
let's consider a macroscopic isolated system
${\cal S}$ at equilibrium, with a total energy density
$\epsilon$, and let $\beta^{-1}_B=\beta^{-1}_{et}$ be
the temperature of the system.
We can now consider a small piece of such a system,
${\cal S}^\prime$, 
which is composed only by few particles.
Obviously, the energy density $\epsilon^\prime$ of ${\cal S}^\prime$
will fluctuate with time. But, we can imagine to isolate
the small system from the rest, exactly when the value
of its energy density is $\epsilon$. Obviously, in this
state, also the energy density of the rest of the system
amount to $\epsilon$. We have two isolated (sub)systems,
a macroscopic one and a small one.
\textit{Now, by using the surface entropy, we obtain the same
estimation of the temperature for both the systems,
${\cal S}$ and ${\cal S}^\prime$.
On the contrary, the Boltzmann entropy 
provides a different estimation for the temperature of
the small system.}

\section{Derivation of the surface entropy \label{derivation}}
We have emphasized that the surface entropy fixes some
of the pathologies that the Boltzmann entropy exhibits
in the statistical description of small systems. 
\textit{By pushing our reasoning forward, we speculate
that the issues shown by the Boltzmann entropy in
the statistical description of small systems is
actually the signature of a deeper flaw
in its definition.}

%\subsection{Microcanonical ensemble}

\begin{figure}[h]
 \includegraphics[width=7.cm]{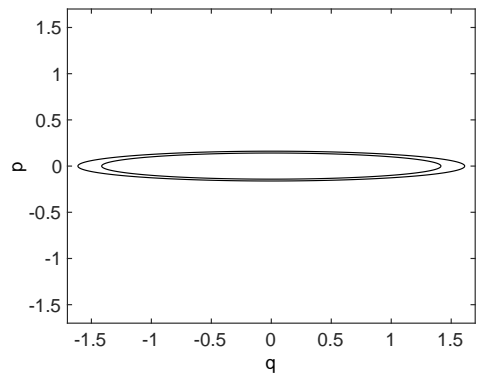}
\caption{ The figure shows two surfaces of constant energy for a
one-dimensional harmonic oscillator with Hamiltonian
$H=\beta^2 p^2/2  + q^2/2$. The two ellipses correspond to the
choice $\beta=10$ and energy values  $E_1 =1$ (internal) and
$E_2=1.3$ (external). The area difference is a strip of
non-uniform thickness representing the inverse of the
gradient norm of $H$, that is the non-uniform density
probability of Eq. \eqref{mie}.
\label{fignew}}
\end{figure}

The postulate of equal a-priori probability, for
a macroscopic system in thermodynamic equilibrium, asserts
that its state is equally likely to be any state satisfying 
the macroscopic conditions of the system. Now, in the case of
the microcanonical ensemble, usually one assumes as
macroscopic condition for the system that its energy has values
within a small interval of amplitude $\Delta E$
around $E$.
Thus, the postulate of equal a-priori probability
leads to the probability distribution 
\begin{equation}
\rho (x) = \left\{
\begin{array}{l}
constant \, , \quad \mathrm{if} \, E\leq H(x) \leq E + \Delta E \, , \\
0 \, , \quad \mathrm{otherwise} \, . 
\end{array}
\right.
\end{equation}
In the limit $\Delta E \to 0$, this density leads to the
expression for the measure on the energy-level
hypersurfaces
\begin{equation}
\dfrac{m^{N-1}(\Sigma_E)}{\Vert\nabla H(x) \Vert} \, ,
\label{mie}
\end{equation}
whence one obtains the Boltzmann entropy.
Nevertheless, if we assume as macroscopic condition
for the system that its energy has a fixed value $E$, 
the postulate of equal a-priori probability leads to the 
probability distribution for the system
\begin{equation}
\rho (x) = \left\{
\begin{array}{l}
constant \, , \quad \mathrm{if} \, H(x) = E \, , \\
0 \, , \quad \mathrm{otherwise} \, , 
\end{array}
\right.
\end{equation}
that is the measure on the energy-level
hypersurfaces
\begin{equation}
{m^{N-1}(\Sigma_E)} \, ,
\label{mni}
\end{equation}
whence the surface entropy definition descends.
Surface and Boltzmann entropies are inequivalent since the first assumes a uniform probability density on the constant energy hypersurface whereas in the case of the Boltzmann entropy the probability density is $1/\Vert \nabla H \Vert$. This entails in the second case a ``concentration'' of probability near to the points where the norm of the Hamiltonian gradient decreases. This is well shown in the simple example reported in Fig. \ref{fignew}.

From this point of view, the two entropies derive from
the two different definitions of isolated system.

For a classical isolated system, one can assume that
the value of its total energy does not range in some
finite interval of values but has some assigned value. 
Also within a quantum description, one can always assume
an isolated system to be in an eigenstate of the
Hamiltonian operator. Moreover, both in the classical
and quantum cases, the energy indetermination
$\Delta E$ is difficult to be
quantified: it should be related
to a measure process, that has a restricted meaning in
the equilibrium theory. In other words, $\Delta E$ seems
not to be a well-defined quantity which one gets rid of
in the limit $\Delta E \to 0$ process.

%A further issue intrinsic to the standard Boltzmann entropy
%definition is the following. By assuming that each state
%of the system with energy in some interval of amplitude
%$\Delta E$  is equally likely, in some sense, it is
%equivalent to assuming that the entropy doesn't depend
%on the value of the energy when it ranges within such
%interval.
%Nevertheless, after that the limit $\Delta E \to 0$ is
%taken, the dependence of the entropy on the value
%of the energy is re-established.    

On the other hand, it is important to emphasize the
following property of the standard Boltzmann entropy.
For the Boltzmann entropy, the measure on phase-space
\eqref{mie} coincides with the microcanonical
invariant measure of the Hamiltonian dynamics.
This fact has two main consequences:
first, the possibility to measure the thermodynamical
average of any physical quantity by means of time averages
along a trajectory (or with phase-space integrals if the
ergodic theorem holds true);
second, and more important, the statistical weight of
each microscopic configuration is invariant under the
Hamiltonian flow.
The first point is not an issue. In fact, the
thermodynamic quantities, as, for instance, temperature
and specific-heat, are defined in term of the derivatives
of the entropy with respect to the energy, and in the
case of the surface entropy, these derivatives can be
measured via microcanonical averages along the
dynamics, by using tricks analogous to that used in
Eq. \eqref{betaSurf}.
As to the second point, the measure \eqref{mni} is not
invariant under the Hamiltonian flux. Nevertheless,
$\Sigma_E$, its volume, and, consistently, $S_s$ are invariant
under the Hamiltonian flux. This means that in the
case of the surface entropy the invariance under the
dynamics is a global property rather than local.
The limitation from local to global of the invariance
under the Hamiltonian flux that we have in the case
of the surface entropy has no physical consequences
since it is not experimentally observable.
In fact, the thermodynamical quantities of an isolated
system at equilibrium depend on the global entropy and
not from local quantities. 
In addition, in order to measure in an experiment a
given local property, it is necessary an interaction
with an external apparatus and this makes the system
no more isolated. 

\section{Consequences of the surface entropy \label{consequences}}
From a geometric point of view, the temperature derived
from the surface entropy has an interesting interpretation:
it is the average of the mean curvature
of $\Sigma_E$ (the hypersurface $H(x) = E$) which is
a geometric quantity.
In fact, in
\cite{Franzosi_PhysicaA18} we have shown that in the case of
the surface entropy it results
\begin{equation}
\beta_s = 
\dfrac{  \int_{\Sigma_E} M(x) m^{N-1}(\Sigma_E) }
{ \int_{\Sigma_E} m^{N-1}(\Sigma_E) } \, , 
\label{meancurvature}
\end{equation}
where
\begin{equation}
M(x) = \dfrac{1}{\Vert \nabla H \Vert}
\nabla \cdot \left(
\dfrac{\nabla H}{\Vert \nabla H \Vert}
\right)
\label{meancurvature1}
\end{equation}
is the local mean curvature divided by $\Vert \nabla H \Vert$.
The latter term makes intensive the quantity measured with
this average, as it is requested for the inverse temperature.

It is worth noting that it might be necessary to use the
surface entropy instead of the usual Boltzmann form, also
in the case of macroscopic systems in which the
phase-space volume scales up as a function of the
number of degrees of freedom of the system 
slower than exponentially. 
This is the case of the long-range interactions.
In fact, in systems with short-range interaction the
measure of the volumes $H(x) \leq E$ concentrates on
the hypersurface $H(x)=E$ when the number of the degrees
of freedoms increases, whereas for systems with long-range
interaction the same is not necessarily true.
Therefore, for the latter systems, the differences
between the quantities derived using the surface and
the Boltzmann entropy can be measurable also up to
macroscopic sizes of the system.

A further domain of application of the surface entropy
may be the problem of negative specific heats
in the microcanonical ensemble \cite{CarignanoEPL2010}.
This phenomenon, that seems to violate the laws of
thermodynamics, takes place in small systems (absolutely
far from the thermodynamic limit), that is in systems
with size smaller or comparable
to the range of the forces driving their dynamics.
For this reason, in these systems the equivalence between
canonical and microcanonical ensembles is not valid and
the microcanonical approach is the only that explains
the emergence of negative specific heat.
Examples of systems of this kind are found in different
fields: we just mention
clusters of sodium atoms \cite{Na},
clusters of stars \cite{stars} and plasma \cite{plasma}.

\begin{acknowledgments}
We are grateful to A. Smerzi, M. Gabbrielli, 
P. Buonsante and especially to Prof. G. Zammori
for useful discussions.

We acknowledge QuantERA support with the project ``Q-Clocks''
and the European Commission.
\end{acknowledgments}

\bibliography{references}

\end{document}